\documentclass[aps,twocolumn,prl,preprintnumbers,amsmath,amssymb,superscriptaddress,nofootinbib]{revtex4}
\usepackage{graphicx,epsfig}
\usepackage{dcolumn}
\usepackage{bm}
\usepackage{color}

\begin{document}

\title{Other-regarding preferences and altruistic punishment:\\ A Darwinian perspective}

\author{Moritz Hetzer}
\affiliation{Chair of Entrepreneurial Risks, Department of
Management, Technology and Economics, ETH-Zurich, CH-8032 Zurich,
Switzerland}

\author{Didier Sornette}

\affiliation{Chair of Entrepreneurial Risks, Department of
Management, Technology and Economics, ETH-Zurich, CH-8032 Zurich,
Switzerland}

\affiliation{Swiss Finance Institute, c/o University of Geneva, 40
blvd. Du Pont dÕArve CH 1211 Geneva 4, Switzerland}

\date{September 4, 2009}



\begin{abstract}

This article examines the effect of different other-regarding
preference types on the emergence of altruistic punishment behavior
from an evolutionary perspective. Our findings corroborate,
complement, and interlink the experimental and theoretical
literature that has shown the importance of other-regarding behavior
in various decision settings. We find that a selfish variant of
inequity aversion is sufficient to quantitatively explain the level
of punishment observed in contemporary experiments: If
disadvantageous inequity aversion is the predominant preference
type, altruistic punishment emerges in our model to a level that
precisely matches the empirical observations. We use a new approach
that closely combines empirical results from a public goods
experiment together with an evolutionary simulation model. Hereby we
apply ideas from behavioral economics, complex system science, and
evolutionary biology.


\end{abstract}


\maketitle

The convergence of individual behaviors to common norms and the
punishment of norm violators is an often observed pattern in groups
of animals and human societies
\cite{Homans1974,Coleman1998,Whiten2005,Bernhard2006,Gurerk2006}.
From small cliques, to the social order in groups and tribes, all
the way to the legal frameworks of countries, punishment is a
widespread mechanism underlying the formation of common norms
\cite{Fehr2002_1,Fehr2004,Henrich2006}. In particular, altruistic
punishment, i.e., the punishment of norm violators at one's own cost
without personal benefit, is frequent in social dilemmas and is
often used to explain the high level of cooperation in humans
\cite{Fehr2000,Fehr2002,Rockenbach2006,Henrich2006,Herrmann2008}.
Within standard economic theory, which relies on rational
selfishness and the dominance of self-regarding preferences, such
behaviors are puzzling, if not disrupting. This observation calls
for the identification of the generative mechanism(s) underlying
altruistic punishment and how its occurrence may be context
dependent.

Laboratory experiments and field studies suggest that egalitarian
motives and other-regarding preferences, which relate a person's
decision to her social environment, have a significant influence in
social dilemmas and in bargaining
\cite{FehrFischbacher2002,Fowler2005,FehrEgalitarianChildren2008,Tomasello2008}.
Several extensions of the standard economic approach provide
descriptions of other-regarding preferences by postulating new terms
in utility functions to account for relative income preferences,
envy, inequality aversion and altruism
\cite{FehrSchmidt1999,Bolton2000,FehrSchmidtTheory2006}. While these
approaches are based on plausible assumptions, their evolutionary
validation remains vague and their quantitative coherence with
empirical data unverified.

There is growing evidence from a variety of studies that pro-social
preferences have emerged in hominids over hundreds and thousands of
years, with deep roots going further back as evidenced from recent
studies on primates
\cite{WaalMonkey2003,WaalMonkey2008,RobinsonGenesBehavior2008,Takahashi2008,Fowler2008}.
The diversity of behavioral traits found in different human cultures
may result from different evolutionary trajectories as well as
distinct relative influence of the cultural versus genetic heritages
\cite{Cason2002,Henrich2006,Hil2004}. A composite picture is
emerging, according to which the perception of fairness, the
reaction to unfair behavior and the individual's response to its
social environment in general, are encoded both in cultural norms
and in genes
\cite{Boyd1988,Laland2000,Gintis2003,Sinha2005,Jablonka2007,Jasny2008,Efferson2008}.

Cultural norms and genes appear to be subjected to complex
coevolutionary processes occurring over a spectrum of different time
scales. Cultural evolution is shaped by biological conditions,
while, simultaneously, genes are altered in response to the
evolutionary forces induced by the cultural context. The
co-evolutionary dynamics and inter-dependencies of genes and
cultural norms constitutes our starting point to understand the
properties of social preferences revealed in experimental economics,
field studies and, of course, in real life.

Experiments on public goods and social dilemma games provide
convenient tools to study social preferences in well-defined
scenarios under controlled conditions. In particular when designed
with the opportunity to punish other subjects at own costs,
altruistic behavior is manifested
\cite{Fehr2000,Fehr2002,Decker2003,Masclet2003,Noussair2005}. In
these experiments, one can study in details what controls the
predisposition of humans to bear the costs associated with
punishments of free riders, and how it may improve the welfare of
the group. Even in one-shot interactions in public good games for
which reputation and reciprocal effects are absent, punishment,
which is costly to the punisher and thus in contradiction with
rational choice theory, is frequently observed
\cite{Fehr2000,Fehr2002,Anderson2006}.

Here we develop computer simulations of synthetic agents within an
agent-based model (ABM), that describes the long-term co-evolution
of norms and genes in populations being exposed to a typical public
goods dilemma. Our work can be viewed as an extension of the
literature on ABM approaches to the evolution of cooperation
\cite{Axelrod1985,Axelrod1997}. Specifically, we set our ABM to
compare with the results of two public goods game experiments
conducted by Fehr and Gachter \cite{Fehr2000,Fehr2002}. Our modeling
strategy is to see the empirical observations in Fehr and Gachter's
experiments as a snapshot within a long-term evolutionary dynamic.
Our ABM mimics the norm-gene co-evolution that has occurred over
hundreds and thousands of years. We calibrate our model by means of
empirical data, to quantitatively identify the underlying preference
types that drive the observed contemporary behavior in the
corresponding dilemmas. In doing so, our goal is to determine the
conditions under which agents develop spontaneously a propensity to
``altruistically'' punish, starting from an initial population of
uncooperative non-punishers. Here, we specifically look into a set
of common assumptions made by economists to account for altruistic
punishment behavior within the framework of utility theory:
Other-regarding preferences in form of inequality and inequity
aversion.

Initialized by variants of these other-regarding preferences, the
traits of our agents converge to statistically stable distributions
after long transients, which are taken to describe the present-day
characteristics of modern humans. In other words, the experiments of
Fehr and Gachter \cite{Fehr2000,Fehr2002} are interpreted as
sampling the statistically stationary characteristics of a cultural
group of subjects\footnote{Here undergraduate students from the
Federal Institute of Technology (ETH) and the University of Zurich}
which have evolved over a long time horizon. Their response to
specific social dilemma situations are then revealed through the
present-day experiments. One should, however, keep in mind that
other patterns of behaviors may have emerged under different norms
and genetic endowments.

\section{Empirical motivation}

The design of our ABM is inspired by the two public goods game
experiments conducted by Fehr and Gachter \cite{Fehr2000,Fehr2002}.
In these experiments, subjects are arranged in groups of $n=4$
persons. At the beginning of each period, subjects received an
initial endowment of 20 monetary units (MUs). Thereafter, subjects
could invest $m \in [0,20]$ MUs to a common group project, which
returned $g_1=1.6$ MUs for each invested MU. The total return from
the project was then equally split and redistributed to all group
members. Thus, the marginal return per capita was $g_1/n=0.4$. As
long as $g_1/n <1$, the game has a vivid social dilemma component
and the setup is susceptible to defection through material
self-interest, since it is rationally optimal not to cooperate,
while the group is better off if each member cooperates. Hence one
can consider the subjects' investment as their level of cooperation.

In a second extended run, subjects were additionally provided with
the opportunity to punish other group members, after they have
received the project return and have been informed about the
individual contributions. The use of punishment was associated with
costs for both parties, in which each MU spent by a punisher led to
$r_p=3$ MUs taken from the punished subject
\cite{Fehr2002}\footnote{In \cite{Fehr2000} the punisher paid
approximately 2 MUs to take an additional 10\% from the punished
subject's period profit.}. The fact that punishment is costly and
that the cost to the punished one is larger ($r_p>1$) are important
properties of the experimental design, which are thought to capture
schematically many real life situations. Versions with and without
punishment were played both in a partner treatment, in which the
group composition did not change across periods, and in a stranger
treatment. In the later, subjects were reassigned to new groups at
each period and thus were only engaged in one-shot interactions
during the entire runtime of the experiment. In total, the
experiments were played for $T_1=10$ \cite{Fehr2000} and $T_2=6$
periods \cite{Fehr2002} respectively\footnote{To avoid the {\it last
round} effect, we consider only data from periods $1-9$ and $1-5$,
respectively.}.

The data from Fehr and Gaechter as well as from several other public
goods experiments \cite{Decker2003,Masclet2003,Noussair2005} show
that people, if provided the opportunity, frequently punish
defectors, even if this is costly to themselves. In the case of
repeated interactions, as in the partner treatment, such behavior
can be explained in a standard way as resulting from a strategic
optimization performed by rational selfish agents, who select the
``direct reciprocity'' mechanism for cooperation. What is more
surprising is that strangers continue to punish at a cost to
themselves even in one-shot interactions for which there is no
positive material gain, even in absence of mechanisms associated
with direct, indirect reciprocity and reputation building. This
behavior is referred to as ``altruistic punishment'' to emphasize
the conflict with the behavior expected from purely rational agents.
The question we address here is why humans behave in a way that
strictly contradicts rational choice, i.e., they continue to
cooperate and punish at a cost to themselves.

\section{The agent-based model}

We extend the setup of the altruistic punishment game of Fehr and
Gachter \cite{Fehr2002} to construct an ABM of a population of
agents who play a public goods game with punishment, while adapting
and evolving over long periods according to generic evolutionary
dynamics. On the short time scales of Fehr and Gachter's
experiments, the traits of the human players probed by the games can
be considered fixed for each player. In contrast, our evolutionary
ABM aims at determining which superordinate regime of
other-regarding preferences have led our ancestors to develop traits
promoting altruistic punishment behavior. These traits might again
be encoded in the cultural context, in genes, or both. Thus, we let
the traits evolve over time according to standard evolutionary
dynamics: Adaptation, selection, crossover and mutation. In order to
capture the possible evolution of the population, agents adapt and
die when unfit. Newborn agents replace the dead ones, with traits
taken from the pool of the other surviving agents. Starting from an
initial population of agents with no cooperation and with no
propensity to punish others, we will find the emergence of long-term
stationary populations whose traits are interpreted to represent
those probed by contemporary experiments, such as those of Fehr and
Gachter.

The results presented below correspond to groups of $n=4$ agents as
in the Fehr and Gachter's experiments. At the beginning of the
simulation (time $t=0$), each agent is endowed with $w_i(0)=0$
MUs\footnote{Only the transient behavior is sensitive to the choice
of this initial wealth while the long-term results are independent
of this initial value.}, which represents its (human and material)
capital. Each agent $(i)$ is characterized by two traits $[m_i(t):
k_i(t)]$, her level of cooperation and her propensity to punish,
that are subjected to evolutionary forces.

\subsection{ABM algorithm}
A given simulation period $t$ is decomposed into five sub-periods
with the following chronology.
\begin{enumerate}
\item {\bf Cooperation}. Each agent $i$ chooses an amount to
contribute to the group project referred to as her level of
cooperation. Combining all the contributions by all group members
and splitting it equally, this leads to a return per agent given by
equation (\ref{eq:project_return}).
\item {\bf Punishment}. Given the returns from the project and the
agents' cooperation levels which are revealed to all, each agent may
choose to punish the other agents according to the rule defined by
equation (\ref{eq:k_punish}).
\item {\bf Consumption}. Each agent consumes the amount defined by
equation (\ref{eq:consumption_share}).
\item {\bf Adaptation}. Given her profit or loss (P\&L) defined by
equation \ref{eq:profit_with_pun} , which results from the project
return minus her contribution and minus the cost of punishments, an
agent may adapt her cooperation level $m_i(t)$ and her propensity to
punish $k_i(t)$.
\item {\bf Evolution}. An agent whose capital drops below $0$
dies and is replaced by another agent whose traits are inherited
from the pool of survival agents with some additional mutation.
\end{enumerate}
These five steps are now described in details.

\subsection{Cooperation step}

As denoted before, the first trait of a given agent is her intrinsic
willingness to cooperate denoted by $m_i(t)$. At each period $t$,
the agent $i$ contributes $m_i(t)$ MUs to the group project. As in
the lab experiment of Fehr and Gachter, each MU invested in the
group project returns $g_1=1.6$ MUs to the group. Each agent
receives the same return
\begin{equation}
r(t)=(g_1/n) \cdot \sum_{j=1}^{n} m_j(t) \label{eq:project_return}
\end{equation} from the group
project, resulting in a first-step profit or loss of
\begin{equation}
s_i(t)=r(t)-m_i(t)= (g_1/n) \cdot \sum_{j=1}^{n} m_j(t) ~-~m_i(t)~,
\label{eq:profit_without_pun}
\end{equation}
for a given agent $i$ equal to the difference between the project
return and its contribution in period $t$.

\begin{figure}
\begin{center}
\includegraphics[width=.4\textwidth]{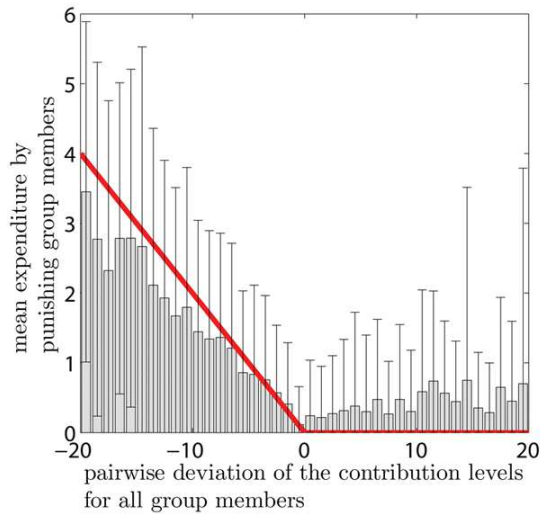}
\end{center}
\caption{Mean expenditure of a given punishing member as a function
of the deviation between her contribution minus that of the punished
member, for all pairs of subjects within a group, as reported
empirically \protect\cite{Fehr2002}. The straight line crossing zero
shows the average decision rule for punishment that our agents
spontaneously evolve to at long times. Its slope $-k \approx -0.2$
defines the average propensity $k$ to punish (see the main text).
The anomalous punishment of cooperators, corresponding to the
positive range along the horizontal axis, is neglected in our
model.}\label{fig:emp_pun_dev}
\end{figure}

\subsection{Punishment step}

Punishments in our ABM follow the same design as in the experiments
of Fehr and Gachter \cite{Fehr2000,Fehr2002}. To choose the agents'
decision rules on when and how much to punish, we are guided by
figure \ref{fig:emp_pun_dev} which shows the mean expenditure of a
given punisher as a function of the deviation between her
contribution minus that of the punished individual, as reported
empirically \cite{Fehr2002}. One can observe an approximate
proportionality between the amount spent for punishing the lesser
contributing agent by the greater contributing agent and the
pairwise difference $m_j(t)-m_i(t)$ of their contributions. The
figure includes data from both the partner and the stranger
treatments in the two sets of experiments \cite{Fehr2000,Fehr2002}.
In our ABM, this linear dependence, with threshold, is chosen to
represent how an agent $i$ decides to punish another agent $j$ by
spending an amount given by
\begin{equation}
p_{i \rightarrow j}(t) =
\begin{cases} k_i(t) \cdot (m_i(t)-m_j(t)) & ~~~~~m_i(t)\ge m_j(t)~,\\
0 & \mbox{otherwise.}
\end{cases}
\label{eq:k_punish}
\end{equation}

The coefficient $k_i(t)$, which represents the propensity to punish,
is the second trait that characterizes agent $i$ at time $t$. It is
not fixed to the average value $k \approx 0.2$ found in the
experiments. It is allowed to vary from agent to agent and it
evolves as a function of the successes and failures experienced by
each agent, as explained below. In addition to being motivated by
the data shown in figure \ref{fig:emp_pun_dev}, the punishment rule
(\ref{eq:k_punish}) can be thought of as a minimalist description of
proportional response to defection. We will see that, given certain
fairness preferences are activated, evolution makes the punishment
propensities $k_i(t)$ self-organize towards a distribution closely
resembling the empirical data.

As a result of being punished the wealth of the punished agent $j$
is reduced by the amount spend by agent $i$ multiplied by the
punishment impact factor $r_p$. As in Fehr and Gachter's second
experiment, we fix the punishment impact factor to $r_p=3$.

\subsection{Consumption step}

Consumption is introduced as a realistic driving force to select for
successful agents and remove unfit ones. At each period, consumption
absorbs an amount $c(t)$ of the capital of each agent. The amount is
assumed to be determined by the social context, specifically as the
change of average group wealth ($\bar{w}(t)$) over the last period:
\begin{equation}
c(t)= {\rm Max}[\bar{w}(t-1)-\bar{w}(t-2)~;~0]~.
\label{eq:consumption_share}
\end{equation}
The proportionality between consumption and income captures the
evidence that, for modern societies, consumption is indeed roughly
proportional to income, at least for $99\%$ of the population that
exclude the super-wealthy \cite{Keynes2006,Mehra2001}. In primitive
societies, larger human and material capital led also to larger
consumption in the form of producing more offsprings\footnote{While
we account for this larger proportional consumption, we do not
include a population dynamics, as our model assumes a constant group
size $n$, with each death being followed by a corresponding birth.}.
The form of consumption (eq. \ref{eq:consumption_share}) is intended
to capture the survival needs and selection pressure that are
determined relative to the social group.

\subsection{Wealth balance}

The total P\&L $\hat{s}_i(t)$ of an agent $i$ over one period of her
life is thus the sum of three components: (i) Her first step P\&L
$s_i(t)$ from the group project (equation
(\ref{eq:profit_without_pun})), (ii) the MUs $\sum_{j \neq i} p_{i
\rightarrow j}(t)$ spent to punish others and (iii) the punishments
$r_p \sum_{j \neq i} p_{j \rightarrow i}(t)$ received from others,
where $p_{i \rightarrow j}(t)$ and $p_{j \rightarrow i}(t)$ are
given by (\ref{eq:k_punish}):
\begin{equation}
\hat{s}_i(t) =   s_i(t) - \sum_{j \neq i} p_{i \rightarrow j}(t) -
r_p \sum_{j \neq i} p_{j \rightarrow i}(t)~.
\label{eq:profit_with_pun}
\end{equation}
Equation \ref{eq:profit_with_pun} represents the second step P\&L of
agent $i$ in period $t$. Putting this all together, the wealth
(fitness) of agent $i$ thus increases or decreases at each period
$t$ according to
\begin{equation}
w_i(t+1)=w_i(t)+\hat{s}_i(t)-c(t)~. \label{prsfdfw}
\end{equation}

\subsection{Adaptation Dynamics}

The traits $[m_i(t); k_i(t)]$ characterizing each agent $i$ at a
given time $t$ evolve with time according to standard evolutionary
dynamics: Adaptation, selection, crossover and mutation. As has been
argued (e.g. by Arthur \cite{Arthur1994} or Holland
\cite{Holland1989}), humans (and our ancestors) are likely to use
inductive reasoning to make decisions. In particular, this means
that humans tend to replace working hypotheses with new ones when
the old ones cease to work. We adopt this bounded rational approach
to define adaptation rules determining the propensity to punish and
the level of cooperations chosen by agents.

{\sf Adaptation of the propensity to punish}. To identify which type
of preference norm drive the evolution of the trait $k_i(t)$
associated with the propensity to punish to a level, that is
observed in the experiments, we test a distinct set of adaption
rules. Each adaption rule corresponds to a specific subset of
other-regarding preference relations. Here, we specifically focus on
different variants of inequality and inequity aversion preferences
and therefore consider the five following types of agents: (A)
self-regarding, (B) inequality averse, (C) inequity averse, (D)
disadvantageous inequality averse and (E) disadvantageous inequity
averse. Here ``disadvantageous'' indicates that agents are only
inequality/inequity averse if the inequality/inequity plays to their
disadvantage, whereas pure inequality or inequity avers agents
dislike both, situations in which they have been discriminated as
well as situations in which they discriminated others.

In each given simulation, we use only homogeneous populations, that
is, we group only agents of the same type.

\begin{itemize}
\item[A:] {\bf self-regarding agents}: In this universe, an agent
updates her propensity to punish only if her P\&L $\hat{s}_i(t)$
given by (\ref{eq:profit_with_pun}) obtained on the previous
investment period turns out to be smaller than her required
consumption $c(t)$. The update consists in an unbiased random
increment according to\footnote{Our results are robust to changes of
the width of the interval, as long as it remains symmetric around
zero.}
\begin{equation}
k_i(t+1)=k_i(t)+\kappa_{[-0.005,0.005]}~, \label{adapt_k}
\end{equation}
where $\kappa$ is a uniformly distributed random number drawn from
the interval indicated in the subscript. Only draws of $\kappa$'s
that ensure positiveness of $k_i(t+1)$ are allowed.

\item[B:] {\bf inequality averse agents}: In this universe,
an agent $i$ updates her propensity to punish if her P\&L
$\hat{s}_i(t)$ given by (\ref{eq:profit_with_pun}) is not within a
specific tolerance range $[-l,+l]$ around the average P\&L of the
other members of her group, i.e. if $(\hat{s}_i(t)<\bar{s}(t)-l)$ or
$(\hat{s}_i(t)>\bar{s}(t)+l)$. When this occurs agent $i$ updates
her $k_i(t)$ according to equation (\ref{adapt_k}). We run multiple
simulations initialized by different values for $l$ as presented in
the results section.

\item[C:] {\bf inequity averse agents}: In this universe,
agents set their P\&L in relation to their contributions: An agent
$i$ updates her propensity to punish according to eq.
(\ref{adapt_k}), if...

\textit{(upside inequity)} ...she has contributed less than or
equally to her group fellows ($m_i(t) \leq \bar{m}(t))$, where the
average $\bar{m}(t)$ is performed over the contributions of the
other members of her group and, at the same time, has received a
total P\&L $\hat{s}_i(t)$ defined in (\ref{eq:profit_with_pun})
larger than or equal to the group average ($\hat{s}_i(t) \ge
\bar{s}(t))$, where the average $\bar{s}(t)$ is performed over the
other group members)...

\textit{(downside inequity)} ...or she has contributed more than or
equally to her group fellows ($m_i(t) \ge \bar{m}(t)$) and, at the
same time, has received a total P\&L less than or equal to the group
average ($\hat{s}_i(t) \leq \bar{s}(t)$).

\item[D:] {\bf disadvantageous inequality averse agents}: In this universe,
agents only dislike situations in which the inequality is to their's
disadvantage:

\textit{(downside inequality)} An agent $i$ updates her propensity
to punish only if her P\&L $\hat{s}_i(t)$ given by
(\ref{eq:profit_with_pun}) is smaller than the average P\&L of the
other members of her group, i.e. ($\hat{s}_i(t)<\bar{s}(t)$). When
this occurs for an agent $i$, she updates her $k_i(t)$ according to
equation (\ref{adapt_k}).

\item[E:] {\bf disadvantageous inequity averse agents}: Likewise to setup
(C), agents set their P\&L in relation to their contributions,
however they only dislike situation in which the inequity is
detrimental to them.

\textit{(downside inequity)} If an agent $i$ has contributed equally
or more than her fellows in the group ($m_i(t) \ge \bar{m}(t)$) and,
at the same time, has received a total P\&L $\hat{s}_i(t)$ defined
in (\ref{eq:profit_with_pun}) smaller than or equal to the group
average ($\hat{s}_i(t) \leq \bar{s}(t)$), then she updates her
propensity to punish according to eq. (\ref{adapt_k}).
\end{itemize}

{\sf Adaptation of the cooperation level}. Following the proverb
``necessity is the mother of all invention,'' agents adapt their
cooperation level $m_i(t)$ under adverse conditions, i.e., when
their P\&L in the last investment round is smaller than the required
consumption $c(t)$. When this adverse situation occurs, an agent
randomly updates her contribution according to
$m_i(t+1)=m_i(t)+\epsilon_{[-0.005,0.005]}$, where $\epsilon$ is a
random number uniformly distributed in the interval indicated in the
subscript.\footnote{Our results are robust to changes of the width
of this interval as long as it remains symmetrical.} Since
contributions are non-negative, only those draws of $\epsilon$ are
kept that ensure the positiveness of $m_i(t+1)$.

\subsection{Replicator Dynamics: Selection, crossover and mutation}

In addition to the adaptation of the agents' traits $[m_i(t);
k_i(t)]$ described above, evolution occurs by replacing
under-performing agents. When an agent's wealth $w_i(t)$ drops below
zero, the agent dies and is replaced by a new one with different
traits $[m_i(t+1), k_i(t+1)]$, determined by those of the surviving
agents of the group. The following variants give essentially the
same results.
\begin{itemize}
\item To capture the phenomenon that norms and genes are
inherited from the pool of existing agents, a first variant assumes
that the new agent has her traits simply equal to the average of
those of the surviving agents.
\item A second variant consists in assuming a noisy
inheritance of the average traits of the group members.
Specifically, the new agent has traits equal to the average of those
of the surviving agents plus a noise proportional to the standard
deviation of the surviving agents' traits.
\item A third variant first determines clusters among the
agents, including the newborn, using a standard clustering algorithm
on the vectors of the two traits. The existence of several clusters
is taken to account for the possible emergence of heterogeneous sets
of norms within the group. Then, the newborn takes the average
traits of the agents in her cluster, decorated by an additional
mutation implemented by adding a random number to each trait
proportional to the standard deviation of the surviving agents'
traits.

\end{itemize}

\section{The emergence of altruistic punishment}

We have run our ABM with thousands of independent groups of $n=4$
agents over one million simulation periods $t$. The agents have a
double peak distribution of lifetimes, with a mean of the order of
$1.7 \cdot 10^4$ periods and a median of the order of $1.5 \cdot
10^4$ periods\footnote{These lifetimes correspond to a population of
disadvantageous inequity avers agents (E)}. This expresses that,
while many agents die at an early age, also many survive beyond
their ``childhood'', enabling them to adapt their traits. Each run
thus corresponds to several hundreds of generations. Each simulation
has been initialized with all agents being uncooperative
non-punishers, i.e., $k_i(0)=0$ and $m_i(0)=0$ for all $i$'s. After
a long transient, we observe that the distribution of propensities
to punish converges to different stationary functions depending on
which adaptation dynamics (A,B,C,D or E) for $k$ is active.

\begin{figure}
\begin{center}
\includegraphics[width=.47\textwidth]{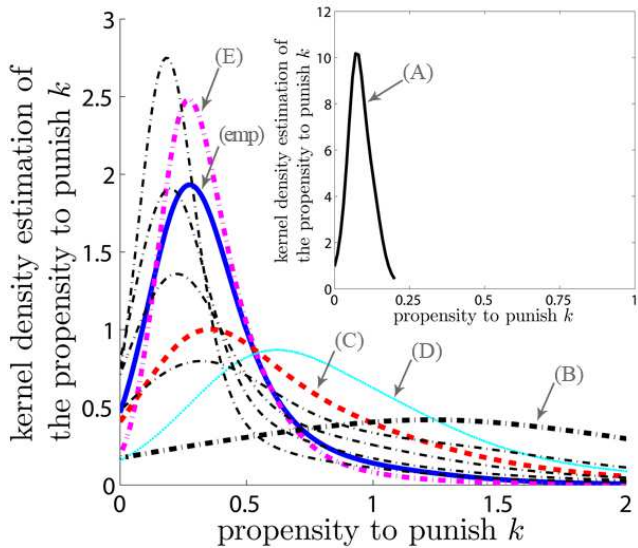}
\caption{ Distributions of propensities to punish in Fehr and
Gachter's experiments (emp) and in our ABM at long times under the
five different adaptation dynamics (A to E) under 800 system
realizations. The inset shows the smoothed distribution for the
self-regarding agent population (adaptation dynamics A), while the
main frame plots the smoothed distributions obtained by the
adaptation rules B to E as well as for the empirical data (emp). All
distributions has been smoothed by a standard kernel method. (emp):
Distribution estimated from Fehr and Gachter's experiments using the
procedure explained in the text; the other distribution estimates
correspond to the adaptation dynamic A to E. For adaptation dynamic
(B) the plot shows the obtained distributions for all tolerance
range parameters $l \in {0.1;0.075;0.05;0.025;0}$ (from the mode
close to the origin ($l=0.1$) to the mode farthest ($l=0$). The
parameters of our ABM simulations are: $n=4, g_1=1.6,
r_p=3$.}\label{fig:sim_k_kernel}
\end{center}
\end{figure}

In the following we give a qualitative analysis of our simulation
results. Selfish agents adapting their propensity to punish
according to dynamics (A) remain weak punishers, and no significant
``altruistic punishment'' is observed as shown in the inset of
figure \ref{fig:sim_k_kernel}. In contrast, for agents endowed with
inequality or inequity aversion (adaptation rules B to E),
stationary states of the propensity to punish emerge spontaneously,
each with different characteristics. For all adaptation rules (B to
E) it holds, that altruistic punishment has emerged endogenously as
an evolutionary stable trait in the competitive resource-limited
world described by our model. The responsible key ingredients are
the variants of other-regarding preferences (B,C,D or E). It should
be stressed that a symmetric (upside and downside)
inequity/inequality aversion is not needed as a condition to let
altruistic punishment emerge. The selfish disadvantageous inequality
or inequity aversion (dynamics D and E) is sufficient.

We now turn to a quantitative characterization of the properties of
the altruistic cooperators that evolved in our ABM. Figure
\ref{fig:sim_k_kernel} compares the stationary distributions of the
propensities to punish obtained with our ABM at long times for the
four adaptation dynamics (B to E) with that obtained in Fehr and
Gachter's experiment\footnote{We used the data from the
\cite{Fehr2000,Fehr2002} stranger treatments.}. The propensities to
punish in the experimental data have been inferred as follows:
Knowing the contributions $m_i >m_j$ of two subjects $i$ and $j$ and
the punishment level $p_{i\rightarrow j}$ of subject $i$ on subject
$j$, the propensity to punish characterizing subject $i$ is
determined by
\begin{equation}
k_i=-\frac{p_{i\rightarrow j}}{m_j-m_i}~.
 \label{eq:k_emp_all}
\end{equation}
Applying this recipe to all pairs of subjects in a given group, we
obtain twelve measures of propensities to punish per group. We then
take the average over these twelve values to obtain a single robust
estimation associated with a given group. Sampling all groups and
all periods, we obtain the distribution shown in figure
\ref{fig:sim_k_kernel} (continuous thick line (emp)).

Figure \ref{fig:sim_k_kernel} suggests that the adaptation dynamics
E (disadvantageous inequity averse agents) is the only one able to
fit the empirical distribution. Indeed, the Kolmogorov-Smirnov test
cannot reject the hypothesis that the $k$ values evolved in our ABM
world for the adaptation dynamics E and the empirical values are
drawn from the same distribution at the 80\% confidence level
($p$-value $\simeq 0.2$). In contrast, the other adaptation dynamics
A to D are strongly rejected ($p$-value $=0$). Given the simplicity
of our ABM and of its underlying assumptions, it is striking to find
such detailed quantitative agreement for one of our dynamics.

For selfish agents (dynamics A), we found that the weak level of
punishment that evolved in the ABM is entirely controlled by
selection pressure, and thus the presence of consumption. In
particular, in the absence of any selection pressure, the level of
altruistic punishment drops to zero. In all other cases, it remains
too small to explain the empirical results of Fehr and Gachter. For
the inequality averse population (B) it become obvious, that within
a symmetric variation of the tolerance range parameter $l$ the
empirical distribution can not be reproduces: Figure
\ref{fig:sim_k_kernel} shows the resulting distributions for the
tested values $l \in [0;0.025;0.05;0.075;0.1]$ as thin dashed lines,
with $l=0.1$ corresponding to the mode close to the origin going
stepwise to $l=0$ with the mode farthest from the origin. For
inequity avers agents (dynamics C) and disadvantageous inequality
averse agents (dynamics D), we find levels of altruistic punishments
that far exceed the empirical evidence. Our quantitative comparison
with Fehr and Gachter's experiments supports the hypothesis that
human subjects are well-described as being disadvantageous inequity
averse, corroborating and complementing previous evidence
\cite{FehrSchmidt1999,Bolton2000,FehrSchmidtTheory2006}. The results
obtained with our ABM simulations suggest that the co-evolution of
norms and genes promoting altruistic punishment have been influenced
by disadvantageous inequity aversion in the presence of simple
inductive strategies, leading to self-sustained co-evolving traits
made robust by reinforcing feedbacks.

The distribution of propensities $k_i$ to punish exhibits a mode
around $k=0.2$, which means that most punishers spend an amount
approximately equal to one-fifth of the experienced differences in
contributions. Note that the value of the mode around $k=0.2$ is
close to the slope of the straight line fitting the empirical data
shown in figure \ref{fig:emp_pun_dev} providing another confirmation
of the explanatory power of our ABM. This most probable value
$k=0.2$ has also been obtained analytically by assuming an
evolutionary optimization of the expected gains with respect to
potential future losses due to punishment \cite{DarcetSornette2008}.

\section{The effect of the propensity to punish on cooperation}

We now demonstrate with our model that punishment is a key
stabilization mechanism for sustaining cooperation. For this, we
need some destabilizing process that tends to destroy cooperation in
the absence of punishment. The experiments of Fehr and Gachter
suggest such a mechanism. A detailed analysis of the
period-by-period decision outcomes made by human subjects shows
evidence of short-term persistence in their updates of cooperation
levels: Previous changes $m_i(t-2) \to m_i(t-1)$ of contributions
that led to larger return from the project ($s_i(t-1)>s_i(t-2)$) are
followed by subsequent updates $m_i(t-1) \to m_i(t)$, with the same
trend: $[m_i(t) - m_i(t-1)]\cdot [m_i(t-1) - m_i(t-2)]>0$). We refer
to this behavior as using a ``trend-following strategy.''

\begin{figure}
\begin{center}
\includegraphics[width=.35\textwidth]{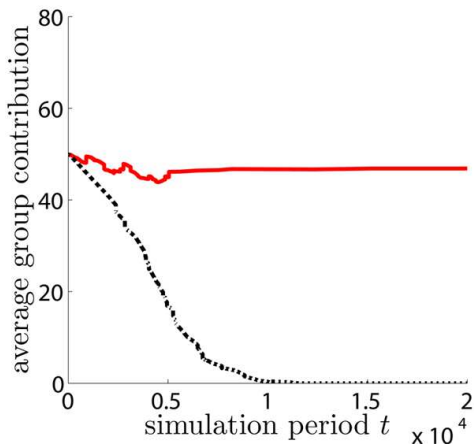}
 \caption{ Average group contribution for a
group of 4 agents with punishment ($k=0.2$ - continuous line) and
without ($k=0$ - dashed line) over 20,000 simulation periods and 16
system realizations. The initial contribution $m_i(0)$ for all
agents $i$ of a group is randomly drawn form a uniform distribution
in $\{49,51\}$.}\label{fig:sim_punEffect}
\end{center}
\end{figure}

When we add the trend-following strategy to our 4-step algorithm as
described above but, in the absence of punishment (all $k$'s are
imposed equal to $0$), we find that cooperation that was maintained
previously in the presence of punishment decays after a few thousand
periods as shown in figure \ref{fig:sim_punEffect}. In contrast, if
punishment is restored in the presence of the destabilizing
trend-following strategy, cooperation remains stable. Note also that
the emergence of the stable distribution of propensity to punish
reported in figure \ref{fig:sim_k_kernel} is robust to the addition
of the trend-following strategies used by the agents.

\begin{figure}
\begin{center}
\includegraphics[width=.35\textwidth]{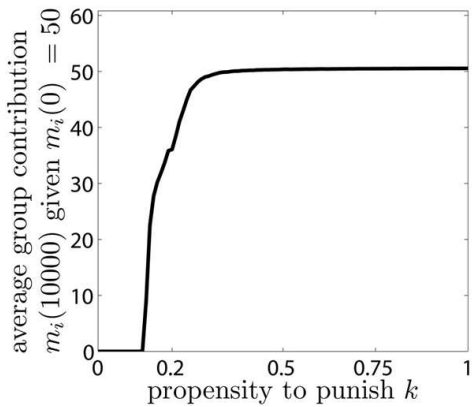}
 \caption{ Average group contribution for a
group of 4 agents as a function of $k$ after an equilibrium time of
20000 simulation periods and 800 system realizations. $k$ is fixed
to the corresponding value on the x-axis and the initial
contribution $m_i(0)$ for all agents $i$ of a group is randomly
drawn form a uniform distribution in
$\{49,51\}$.}\label{fig:sim_punEffect2}
\end{center}
\end{figure}

Figure \ref{fig:sim_punEffect2} shows the average level of
cooperation in a group of 4 agents after a transient period of
20,000 simulation periods for 800 system realizations as a function
of the propensity $k$ to punish, when the level of cooperation for
all agents is initially drawn from a uniformly distributed random
variable in $\{49,51\}$. It reveals that the level of cooperation
undergoes a bifurcation from zero to significant levels of
cooperation, for a value of $k$ close to the mode of the empirical
distribution of the propensities to punish. This suggests that
evolution may have selected an ``optimal'' propensity to punish
altruistically defectors in order to sustain cooperation.

\begin{figure}
\begin{center}
\includegraphics[width=.35\textwidth]{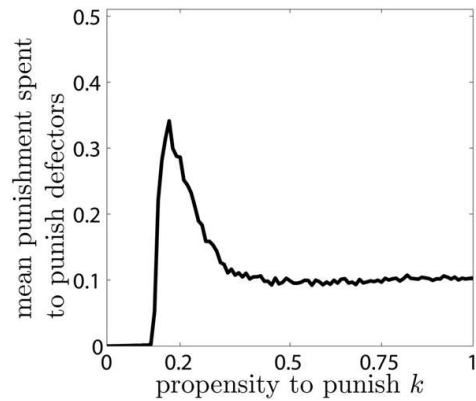}
 \caption{ Average punishment spent to punish defectors for a
group of 4 agents as a function of $k$ after an equilibrium time of
20000 simulation periods and 800 system realizations. $k$ is fixed
to the corresponding value on the x-axis and the initial
contribution $m_i(0)$ in period $0$ for all agents $i$ of a group is
randomly drawn form a uniform distribution in
$\{49,51\}$.}\label{fig:sim_punSpent}
\end{center}
\end{figure}

To corroborate this hypothesis, figure \ref{fig:sim_punSpent} plots
the average amount of MUs spent to punish a defector during 10000
simulation periods for 3200 system realizations as a function of the
propensity to punish $k$. As in the setup of figure
\ref{fig:sim_punEffect2}, the level of cooperation $m_i(t)$ for all
agents is initialized at period $t=0$ by a random variable uniformly
distributed in $\{49,50\}$. We now consider the intrinsic propensity
to punish $k$ as the potential exposure of defectors to being
punished. The results show clearly, that a higher deterrence, i.e. a
higher value of $k$, effectively causes less exertion of costly
punishment in order to maintain a certain level of cooperation and
norm conformity, respectively. This responsive behavior has been
manifested in many empirical observations
\cite{Kleiman2009_2,Kennedy2008}.

Given, that disadvantageous inequity aversion is the dominant
preference type, cooperation and norm conformity can be maintained
at lower costs if the propensity to punish, i.e. the level of
deterrence, exceeds the tipping point of $k$ at $\simeq0.2$. This
again substantiates, that evolution may have selected an ``optimal''
propensity to punish to sustain cooperation and prevent defection in
contexts where people behave disadvantageous inequity averse.
Comparable results have been obtain using a different simulation
model as has been reported in \cite{Kleiman2009}.

\section{Conclusion}

Our first principal result is that a high level of altruistic
punishment behavior emerges spontaneously from a population of
agents who are initially uncooperative and non-punishers. We have
shown how this results from evolution with adaptation, selection,
crossover and mutation, in a population of agents endowed with
different variants of inequality or inequity aversion. We stress
that our use of the terms ``inequality'' or ``inequity aversion''
does not mean that a new term is added in the agents' utility
function that controls their decisions; in contrast, we only assume
that agents may punish as a reaction to differences in observed
contributions to a group project. Thus, a key ingredient is the
possibility for agents to punish, at a cost to themselves, and that
the punishment be efficient (in the sense $r_p>1$). As a consequence
of the co-evolution of traits characterizing agents' cooperation and
punishment, we obtain the emergence of altruistic punishment
behavior. A fully symmetric (upside and downside)
inequity/inequality aversion is not needed to obtain our results,
with the selfish disadvantageous inequality or inequity aversion
being sufficient.

Our second main result is the identification of disadvantageous
inequity aversion as the most relevant underlying mechanism to
explain the emergence and the degree of altruistic punishment
observed in public goods experiments. This result has been obtained
by combining empirical data with an evolutionary agent-based
simulation model in an innovative way. Our ABM is able to reproduce
quantitatively, without adjustable parameters, the experimental
results concerning the most likely levels of punishment behavior, as
well as their full distribution. This result is of particular
importance to substantiate the assumptions made by economists in
order to describe realistic behavior within the framework of
rational choice: Humans exhibit other-regarding, and in particular,
disadvantageous inequity aversion preferences in their decision
process.

In conclusion, we believe that the combination of empirical research
and agent-based modeling as done here can provide deeper insights
into the apparently non-rational behavior of humans. For instance,
with regard to the often-cited importance of altruistic punishment
in promoting cooperation, our ABM provides a flexible and powerful
methodology to answer many remaining research questions, including
the influence of group interactions, competitions between variants
of other-regarding preferences or the interplay of other mechanisms,
as well as more realistic set-ups in which agents are playing
several games simultaneously so as to mimic a real life situation in
which cognitive abilities and human capital are scarce resources.

\begin{acknowledgments}
We are grateful to E. Fehr and S. Gachter for sharing their
unpublished data with us and to U. Hoffrage for constructive remarks
on the manuscript. The work has been partially supported by ZKB
(Z\"urcher Kantonal Bank). We also acknowledge financial support
from the ETH Competence Center ``Coping with Crises in Complex
Socio-Economic Systems" (CCSS) through ETH Research Grant
CH1-01-08-2.
\end{acknowledgments}


\end{document}